\documentclass[sigconf,nonacm]{acmart}
\usepackage{graphicx}
\usepackage{booktabs}
\usepackage{array}
\usepackage{tabularx}
\usepackage{amsmath}

\usepackage{amssymb}
\usepackage{xcolor}
\usepackage{microtype}
\usepackage{float}
\usepackage{subcaption}

\title{LENS: A Staged Design for Interaction Granularity\\in Sequential CTR Prediction}

\author{Yuan Wang, Yue Liu, Jun Zhang, Jie Jiang}
\affiliation{\institution{Tencent Inc., Beijing, China}\country{}}
\email{{leoyuanwang, herculesliu, neoxzhang, zeus}@tencent.com}

\date{}

\begin{document}

\begin{abstract}
In sequential CTR prediction, a central design question is at what granularity the target should interact with the user behaviour sequence. Existing models mainly follow two routes. Raw-item architectures such as DIN let the target score each item in the sequence directly. This relies on well-trained item embeddings and becomes brittle for sparse items. Latent-query architectures such as HyFormer, MixFormer, and OneTrans build query representations by combining the target with other information. This is more robust across item-density regimes but blunter: target-specific control is diluted. We propose \textbf{LENS} to restore target-specific control within these coarser bottlenecks. LENS has two modules: a Target-Conditioned Query Gate (TCQG) for query activation and a Target-Conditioned Position Bias (TCPB) for history retrieval. We further introduce Query-Specific Position Bias (QueryPos), a simple static position-aware reference for latent-query backbones. Across three representative latent-query backbones and four datasets, the combined QueryPos+LENS design achieves positive total-gain point estimates in all twelve evaluated backbone--dataset cells. We also identify a density-dependent conditioning rule: as item density decreases, the optimal condition source shifts from item-only to item-plus-sequence.

\end{abstract}



\maketitle

\section{Introduction}

In sequential CTR prediction, the central design question is not only which architecture to use, but at what granularity the target should interact with the user history. At a coarse level, raw-item architectures let the target score history directly, while latent-query architectures mediate this interaction through learned query representations. We refer to this design axis as \emph{interaction granularity}. Our starting point is that interaction granularity should be treated as an explicit design dimension rather than as a fixed side effect of architecture choice.

\begin{figure}[t]
  \centering
  \includegraphics[width=\linewidth]{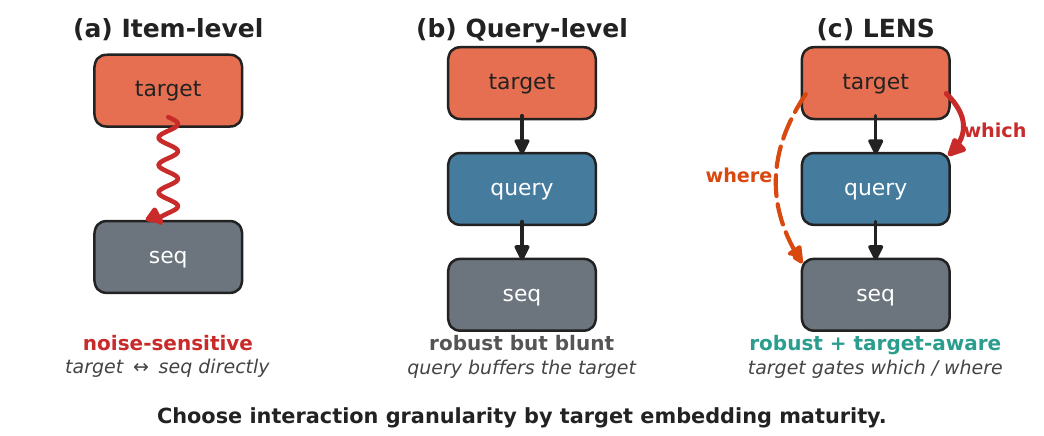}
  \caption{Interaction granularity in sequential CTR. Raw-item architectures (DIN) let the target directly score each history token. Latent-query architectures (HyFormer, MixFormer, OneTrans) mediate target-history interaction through latent query representations. LENS adds target-conditioned residuals that restore target-specific control within these coarser bottlenecks.}
  \label{fig:interaction_granularity}
\end{figure}

At one extreme, DIN~\citep{Zhou2018} attends from the target to each history token. This fine-grained interaction depends on reliable item embeddings, so matching quality degrades on long-tail items with sparse training signal. At the other extreme, latent-query architectures---HyFormer~\citep{HyFormer2025}, MixFormer~\citep{MixFormer2026}, OneTrans~\citep{Chen2025}---build query representations by combining the target with other information. This is more robust across density regimes but blunter: the target's influence is diluted, and the resulting queries are less target-specific (Figure~\ref{fig:interaction_granularity}). Existing backbones therefore expose different interaction granularities, but they do not make that granularity easy to design or enrich explicitly once the architecture is fixed.

We address this gap with a staged design view of target--history interaction for latent-query backbones. Stage~1 selects the interaction bottleneck used by the backbone. Stage~2 adds a position-aware reference through a static position prior. Stage~3 adds target-conditioned residuals that recover target-specific control without changing the backbone's basic computation pattern. Under this view, interaction granularity is not a monolithic architecture decision inside the latent-query family; it can be decomposed into separable design layers that can be analysed and validated incrementally.

We instantiate Stage~2 with \textbf{Query-Specific Position Bias} (QueryPos), a per-query learnable position prior that independently improves latent-query backbones. We instantiate Stage~3 with \textbf{LENS}, a pair of zero-initialised residual modules: a \textbf{Target-Conditioned Query Gate} (TCQG) that controls which latent queries are active for a given candidate, and a \textbf{Target-Conditioned Position Bias} (TCPB) that controls where those active queries retrieve from history. Zero initialisation guarantees that the model starts as the unmodified backbone and can be added incrementally. The primary evidence for this staged view is portability: across four datasets spanning dense to extremely sparse item regimes ($\sim 10^3$ to $\sim 1.1$ samples per item), the four HyFormer cells in the staged validation and the eight external portability cells on MixFormer and OneTrans all show positive total-gain point estimates. This result suggests that the same Stage~2 + Stage~3 additions transfer across multiple latent-query backbones, supporting a portable design recipe within this family.

Beyond the modules themselves, we identify a density-dependent mechanism governing how target conditioning should be formed. The optimal condition source shifts from item-only to item$+$sequence as item density decreases, with a crossover near ${\sim}50$ samples per item. This reveals that the right interaction granularity is not fixed, but adapts to the data regime.

The contributions are: (1) we use \emph{interaction granularity} as an organizing perspective for latent-query sequential CTR models and show that target--history interaction can be analysed as a staged design problem rather than treated only as a fixed consequence of backbone choice; (2) we introduce QueryPos, a position prior that provides a stable position-aware reference across latent-query backbones; (3) we introduce LENS, a pair of target-conditioned query-routing modules that restore target-specific control on top of fixed latent-query bottlenecks, and validate them across three backbones and four density regimes; and (4) we identify a density-driven condition-source rule showing when target-only conditioning is sufficient and when sequence context should be added.

\section{Background}
\label{sec:background}

\subsection{Problem Setup}
\label{sec:full_side}

Each training sample contains a user, a candidate item, a binary click label, a behaviour sequence, and categorical/dense metadata fields (user attributes, item metadata, context). All experiments use the full-side feature protocol: all available non-sequential fields are enabled by default.

\begin{equation}
\hat{y}_{u,i}=f_{\Theta}\!\left(i,\mathcal{S}_u,\mathcal{X}_{u,i}\right)
= P_{\Theta}\!\left(y_{u,i}=1 \mid i,\mathcal{S}_u,\mathcal{X}_{u,i}\right),
\label{eq:problem_setup}
\end{equation}
where $\mathcal{S}_u$ denotes the user behaviour sequence and $\mathcal{X}_{u,i}$ denotes the non-sequential fields for the user--candidate pair and context.

\subsection{Latent-Query Backbones and the HyFormer Reference}
\label{sec:query_decoding}

We use HyFormer as the reference latent-query backbone for staged validation and later test portability to MixFormer and OneTrans. HyFormer-style models~\citep{HyFormer2025} aggregate behaviour sequences through cross-attention from a small set of latent queries to sequence key/value embeddings. With $q$ latent queries and sequence length $L$, the cost is $O(qL)$ rather than $O(L^2)$. The basic HyFormer block alternates \emph{Query Decoding} (cross-attention over the sequence) and \emph{Query Boosting} (token mixing among queries and optionally non-sequential feature tokens). Queries are initialised by a QueryGen projection from the full-side feature vector.

\subsection{Interaction Granularity}
\label{sec:granularity_spectrum}

We define \emph{interaction granularity} as the level at which the candidate modulates history aggregation. This is a design dimension that can be separated from many other backbone choices: an architecture determines an initial interaction bottleneck, but position structure and target-conditioned control need not remain implicit consequences of that bottleneck. In this paper, we only distinguish two coarse regimes: raw-item interaction (item$\times$seq) and latent-query interaction (query$\times$seq). The table below summarises four representative architectures under this view.

\begin{center}
\small
\setlength{\tabcolsep}{3pt}
\resizebox{\columnwidth}{!}{%
\begin{tabular}{lll}
\toprule
\textbf{Backbone} & \textbf{Interaction} & \textbf{Target Entry Point} \\
\midrule
DIN~\citep{Zhou2018} & Raw-item (item$\times$seq) & Direct attn weight per history item \\
HyFormer~\citep{HyFormer2025} & Latent-query (query$\times$seq) & Full-side proj. $\to$ latent queries \\
MixFormer~\citep{MixFormer2026} & Latent-query (query$\times$seq) & Per-head dense proj. incl.\ target \\
OneTrans~\citep{Chen2025} & Latent-query (query$\times$seq) & Single token in unified stream \\
\bottomrule
\end{tabular}
}%
\end{center}

The key limitation is not that these bottlenecks are wrong, but that they are largely fixed by architecture once the backbone is chosen. DIN grants the candidate direct per-item influence. HyFormer, MixFormer, and OneTrans instead mediate the candidate through compressed query representations. Despite their different entry points, these latent-query backbones do not expose dedicated, separable controls for two aspects of candidate-specific behaviour modelling: which query representations should be active for the current candidate, and where those active representations should retrieve from the history. Our staged design in Section~\ref{sec:method} treats these missing controls as separable layers rather than as architecture-specific details.

\section{Related Work}
\label{sec:related_work}

\paragraph{Sequential and Target-Aware CTR.}
Sequential recommendation has progressed through RNN-based models (GRU4Rec~\citep{Hidasi2016}), convolutional approaches (Caser~\citep{Tang2018}), and transformer architectures: SASRec~\citep{Kang2018} adopted causal self-attention, and BERT4Rec~\citep{Sun2019} extended this with bidirectional masked-token training. These backbones are target-agnostic at the encoding stage. DIN~\citep{Zhou2018} introduced target-aware attention where the candidate queries each historical interaction. Extensions include DIEN~\citep{Zhou2019} (interest evolution via GRU), DSIN~\citep{Feng2019DSIN} (session structure), and BST~\citep{Chen2019BST} (Transformer encoder over the sequence). Lifelong-history approaches scaled this paradigm: MIMN~\citep{Pi2019}, SIM~\citep{Pi2020}, ETA~\citep{Chen2021ETA}, and SDIM~\citep{Cao2022SDIM}. A common thread is that target awareness operates at individual history items. LENS instead operates at the query level, providing target specificity without competing with item-level representations.

\paragraph{Latent-Query Architectures and Position Mechanisms.}
The query-decoding paradigm compresses variable-length inputs through learned queries. Perceiver~\citep{Jaegle2021} demonstrated iterative cross-attention from a latent array; Set Transformer~\citep{Lee2019} introduced inducing points; Slot Attention~\citep{Locatello2020} applied iterative slot refinement. In recommendation, MIND~\citep{Li2019MIND} and ComiRec~\citep{Cen2020ComiRec} extract multiple interests but are target-agnostic retrieval encoders. For CTR ranking, HyFormer~\citep{HyFormer2025} uses latent queries with cross-attention and token mixing; MixFormer~\citep{MixFormer2026} co-scales dense interaction and sequence modelling with per-head target projections; OneTrans~\citep{Chen2025} tokenizes all inputs into a single Transformer stream. Although these backbones differ in where the candidate enters the computation, the resulting interaction bottleneck is effectively fixed once the architecture is chosen. LENS extends them by adding explicit target-conditioned controls not present in these architectures. On position encoding, approaches have evolved from sinusoidal encodings~\citep{Vaswani2017Attention} through relative representations~\citep{Shaw2018}, T5 bucketed biases~\citep{Raffel2020}, ALiBi~\citep{Press2022}, RoPE~\citep{Su2024}, and Transformer-XL~\citep{Dai2019}. In CTR, position is typically injected via learned embeddings added to sequence tokens~\citep{Kang2018,Sun2019}. QueryPos differs in being per-query and applied in cross-attention rather than self-attention; TCPB further generates a candidate-specific position residual.

\paragraph{Conditional Modulation and Feature Interaction.}
Feature-wise modulation includes FiLM~\citep{Perez2018}, Squeeze-and-Excitation~\citep{Hu2018}, GLU variants~\citep{Dauphin2017,Shazeer2020}, and HyperNetworks~\citep{Ha2017}. TCQG resembles SE-Net's channel recalibration but conditions on the target-item embedding rather than global context, and the gated representations are latent queries that subsequently attend over the behaviour sequence. Zero initialisation preserves the backbone as a strict subcase. Feature interaction methods---Wide\&Deep~\citep{Cheng2016WideDeep}, DeepFM~\citep{Guo2017DeepFM}, DCN~\citep{Wang2017DCN,Wang2021DCNV2}, AutoInt~\citep{Song2019}, DLRM~\citep{Naumov2019}---are orthogonal to sequence aggregation. On action types, HSTU~\citep{Zhai2024} models heterogeneous actions in a trillion-parameter generative framework. Our typed-exposure approach is simpler: additive action-type embeddings at the input level, requiring no architectural changes, motivated by the embedding-coverage problem in sparse item-vocabulary settings.

\section{Method: LENS as a Staged Design Framework}
\label{sec:method}

\begin{figure*}[!t]
  \centering
  \includegraphics[width=0.85\textwidth]{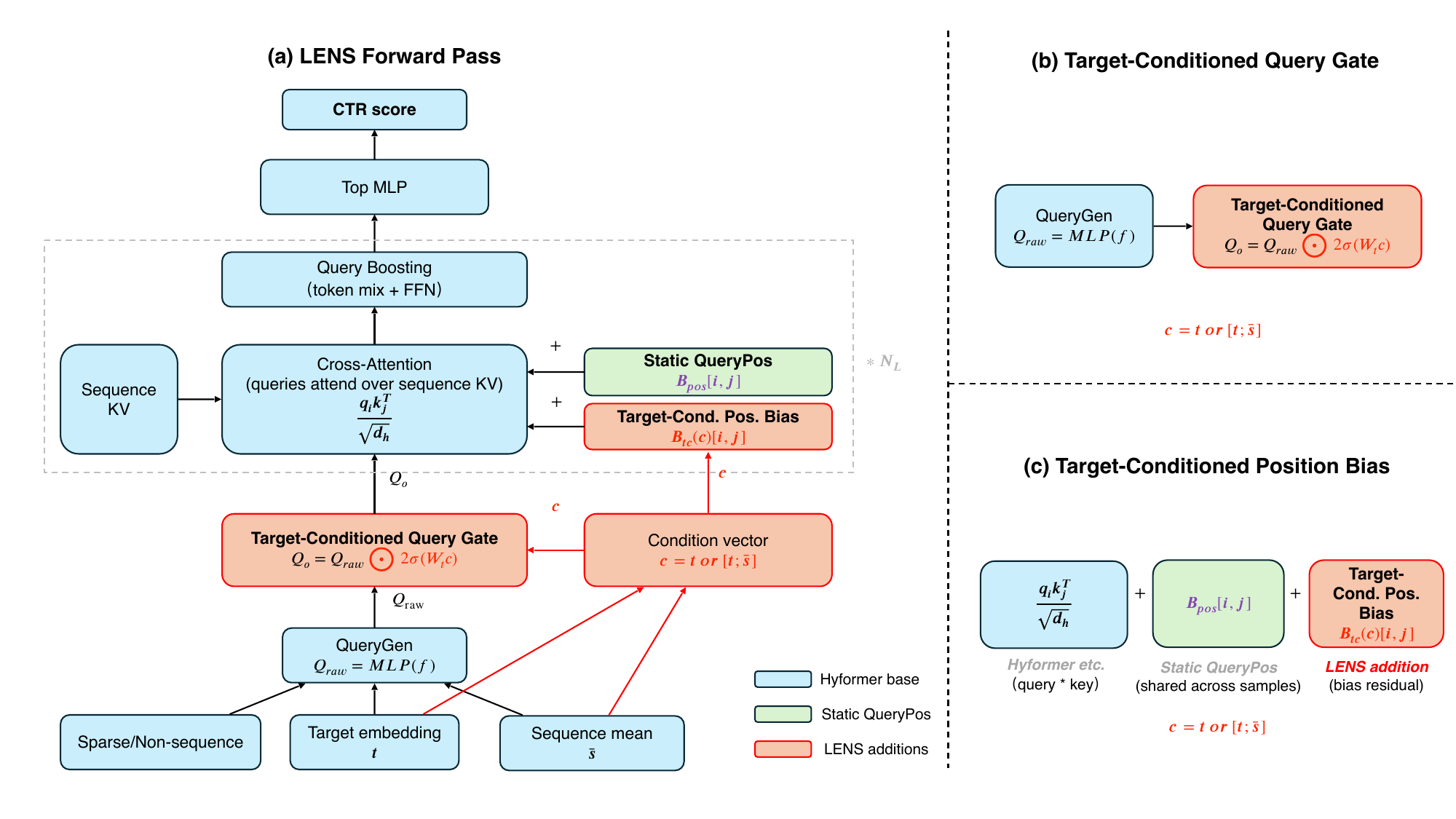}
  \caption{LENS architecture. Colour code follows our staged design: blue denotes Stage~1 (the HyFormer backbone bottleneck), green denotes Stage~2 (the static Query-Specific Position Bias (QueryPos) reference), and red denotes Stage~3 (LENS target-conditioned query routing). \textbf{(a)} Forward pass: QueryGen creates queries from the full-side feature $\mathbf{f}$; queries cross-attend over the sequence across $N_L$ layers with QueryPos and TCPB biases; then boost and read out a CTR score. \textbf{(b)} Target-Conditioned Query Gate: TCQG element-wise multiplies $\mathbf{Q}_{\mathrm{raw}}$ by $2\sigma(\mathbf{W}_t \mathbf{c})$ to yield candidate-specific $\mathbf{Q}_0$. \textbf{(c)} Cross-attention logits decomposition into content (blue), static QueryPos (green), and target-conditioned residual (red).}
  \label{fig:arch}
\end{figure*}

We develop our method as a three-stage design framework for target--history interaction. Stage~1 selects the interaction bottleneck used by the backbone. Stage~2 establishes a position-aware reference that is compatible with that bottleneck. Stage~3 adds LENS target-conditioned query routing to restore candidate-specific control without changing the backbone's basic computation pattern. Figure~\ref{fig:arch} visualizes this staged decomposition directly: blue for the Stage~1 backbone bottleneck, green for the Stage~2 static position-aware reference, and red for the Stage~3 target-conditioned routing. We instantiate the framework on HyFormer as the reference backbone and evaluate its portability to MixFormer and OneTrans in Section~\ref{sec:portability}.

Each stage addresses an orthogonal design dimension: bottleneck selection, static position structure, and target-conditioned modulation. This decomposition lets the gains from different design choices be validated independently rather than conflated inside a single architecture change.

\subsection{Stage 1: Selecting the Interaction Bottleneck}
\label{sec:stage1}

The backbone choice fixes the structural granularity at which target--history interaction occurs. For a query-decoding backbone such as HyFormer, let $\mathbf{t} \in \mathbb{R}^D$ denote the target item embedding, $\mathbf{f} \in \mathbb{R}^{F}$ the concatenated full-side feature representation, and $\mathbf{S} \in \mathbb{R}^{L \times D}$ the behaviour-sequence key/value embeddings. The decoder represents history with $q$ latent queries:
\[
\mathbf{Q}_0 = \operatorname{QueryGen}(\mathbf{f}) \in \mathbb{R}^{q \times D}.
\]
Each layer $\ell$ cross-attends from the latent queries to the behaviour sequence and applies query-token mixing:
\begin{align}
\mathbf{Q}_\ell &\leftarrow \mathbf{Q}_\ell + \operatorname{CrossAttn}_\ell(\operatorname{LN}(\mathbf{Q}_\ell),\; \mathbf{S},\; \mathbf{S}), \\
\mathbf{Q}_\ell &\leftarrow \mathbf{Q}_\ell + \operatorname{Boosting}_\ell(\operatorname{LN}(\mathbf{Q}_\ell),\; \mathbf{X}_{\mathrm{ns}}).
\end{align}

HyFormer exposes three optional capacity switches that define the Stage~1 backbone-selection space in Table~\ref{tab:unified_results}: Seq Pooling Tokens, NS Tokens in Query Boosting, and Per-query FFN. We compare these individual switches, along with the Full Switches combination, and adopt the Seq Pooling Tokens variant as the downstream reference for its minimal cross-dataset variance. Once the backbone is fixed, QueryPos and LENS become portable.

\subsection{Stage 2: Position-Aware Reference}
\label{sec:querypos}

The query decoder is efficient but does not prescribe a position-aware cross-attention design, even though the recency of a behaviour in the history is informative for CTR. QueryPos addresses this gap by injecting a static, candidate-independent position bias into the cross-attention logits.

We evaluate three candidate mechanisms: \emph{Global Position Bias}, which adds a single learnable recency curve shared across all latent queries; \emph{Absolute Position Embedding}, which adds a learned position embedding to each sequence token before key/value projection; and \emph{Query-Specific Position Bias} (QueryPos), which gives each query its own learnable position curve. Part~II of Table~\ref{tab:unified_results} reports the comparison: Global Pos Bias is simple and competitive, Abs Pos Emb is strong on KuaiRec and TAAC but collapses on KuaiRand, and QueryPos stays within $0.0021$ AUC of the per-dataset optimum while maintaining the lowest cross-dataset variance. We adopt QueryPos as the position-aware reference.

QueryPos adds a learnable per-query, per-position prior to cross-attention logits:
\begin{equation}
\operatorname{AttnLogits}(\mathbf{q}_i, \mathbf{k}_j)
= \frac{\mathbf{q}_i \mathbf{k}_j^\top}{\sqrt{d_h}}
  + B_{\mathrm{pos}}^{(\ell)}[i, j],
\label{eq:querypos}
\end{equation}
where $B_{\mathrm{pos}}^{(\ell)} \in \mathbb{R}^{q \times L_{\max}}$ is a layer-specific learnable bias, zero-initialised. The bias is indexed over right-aligned sequence positions: the last valid index corresponds to the most recent behaviour. Each row of $B_{\mathrm{pos}}$ is therefore a position-preference curve for one latent query. Unlike a shared ALiBi-style bias~\citep{Press2022}, each query learns its own curve, reflecting the assumption that different interest dimensions can favour different sequence-index regions: brand preference is often recent, whereas broader category affinity may span a longer prefix.

QueryPos requires only a query/key cross-attention pathway and is therefore backbone-agnostic. We verify in Section~\ref{sec:portability} that adding QueryPos to MixFormer and OneTrans---neither of which includes a dedicated query-specific position component---yields immediate gains without any target conditioning. This shows that static position awareness is a meaningful addition to these backbones, not merely a stepping stone to the target-conditioned modules introduced next.

\subsection{Stage 3: Target-Conditioned Query Routing (LENS)}
\label{sec:lens}

The backbone with QueryPos produces a position-aware query decoder whose behaviour is identical for every candidate item: the same queries attend with the same position biases regardless of what is being scored. LENS introduces candidate specificity by adding two zero-initialised residual modules conditioned on a single vector $\mathbf{c}$. By default $\mathbf{c} = \mathbf{t}$ (target embedding); for sparse target regimes $\mathbf{c} = [\mathbf{t}; \bar{\mathbf{s}}]$, where $\bar{\mathbf{s}}$ is the masked mean of the raw sequence-item embeddings (averaging over non-padding positions), computed from the shared embedding table before projection into the key/value pathway.

The \emph{Target-Conditioned Query Gate} (TCQG) modulates which latent queries are active for the current candidate. The \emph{Target-Conditioned Position Bias} (TCPB) modulates where those active queries retrieve information from the indexed history. The two modules are symmetric: $\mathbf{c}$ controls query activation in TCQG and position retrieval in TCPB.

\subsubsection{Target-Conditioned Query Gate (TCQG).}
\label{sec:tcqg}

The gate replaces the raw query initialisation with a condition-modulated query tensor:
\begin{align}
\mathbf{Q}_{\mathrm{raw}} &= \operatorname{QueryGen}(\mathbf{f}), \\
\mathbf{g}(\mathbf{c}) &= 2\,\sigma\!\left(\operatorname{reshape}(\mathbf{W}_t \mathbf{c},\; q \times D)\right), \label{eq:tcqg_gate} \\
\mathbf{Q}_0 &= \mathbf{Q}_{\mathrm{raw}} \odot \mathbf{g}(\mathbf{c}),
\end{align}
where $\mathbf{W}_t \in \mathbb{R}^{qD \times d_c}$ is zero-initialised and $d_c$ is the condition dimension. At initialisation, $\mathbf{W}_t \mathbf{c} = \mathbf{0}$ and $\mathbf{g}(\mathbf{c}) = 2\sigma(\mathbf{0}) = \mathbf{1}$, so the model reduces exactly to the QueryPos reference. During training, the condition learns to amplify or suppress individual query dimensions, producing a candidate-specific query-activation profile. The factor of $2$ is a centring convention: a plain sigmoid starts the gate at $0.5$ and shrinks the query magnitude, whereas $2\sigma(\cdot)$ preserves the reference scale.

Unlike DIN, which achieves target specificity at the \emph{item level} by attending from the target to each behaviour item, TCQG operates at the \emph{query level}: it modulates which abstract interest dimensions are active before the sequence is aggregated.

\subsubsection{Target-Conditioned Position Bias (TCPB).}
\label{sec:tcpb}

QueryPos is query-specific but static: after training, the same position curve is used for every candidate item. TCPB makes this position-bias residual a function of the target condition. For each layer $\ell$, we project $\mathbf{c}$ into a low-rank query-position mixture:
\begin{align}
\mathbf{M}^{(\ell)}(\mathbf{c}) &=
\operatorname{reshape}(\mathbf{W}_{\tau}^{(\ell)}\mathbf{c},\; q \times r), \label{eq:tcpb_M}\\
\mathbf{P}^{(\ell)} &= \operatorname{PosEmb}^{(\ell)}(1{:}L) \in \mathbb{R}^{L \times r}, \label{eq:tcpb_P}\\
B_{\mathrm{tc}}^{(\ell)}(\mathbf{c})[i,j] &= \langle \mathbf{M}^{(\ell)}(\mathbf{c})_{i,:},\; \mathbf{P}^{(\ell)}_{j,:} \rangle. \label{eq:tcpb_bias}
\end{align}
The resulting $B_{\mathrm{tc}}^{(\ell)}(\mathbf{c}) \in \mathbb{R}^{q \times L}$ is added to the cross-attention logits alongside the static QueryPos bias:
\begin{equation}
\operatorname{AttnLogits}(\mathbf{q}_i, \mathbf{k}_j)
= \frac{\mathbf{q}_i \mathbf{k}_j^\top}{\sqrt{d_h}}
  + B_{\mathrm{pos}}^{(\ell)}[i,j]
  + B_{\mathrm{tc}}^{(\ell)}(\mathbf{c})[i,j].
\label{eq:full_logits}
\end{equation}
We use rank $r=8$. Each layer's condition projection $\mathbf{W}_{\tau}^{(\ell)}$ is zero-initialised, so $B_{\mathrm{tc}}^{(\ell)}(\mathbf{c}) = 0$ at initialisation and the model again reduces to the QueryPos reference. The condition can therefore steer a latent query toward recent-index, mid-history, or longer-range positions on a per-candidate basis, while the static QueryPos bias provides the shared positional scaffold.

\subsubsection{Combined Effect, Initialisation, and Condition Source.}
\label{sec:combined}

TCQG selects \emph{which} queries are active; TCPB steers \emph{where} those queries read. All LENS projections are zero-initialised, so the model starts from the QueryPos reference and is compatible with warm-starting from a pre-trained checkpoint.

The choice of condition source $\mathbf{c}$ is determined by the data regime rather than by model capacity. When the number of training samples per item exceeds approximately 50, the target ID embedding has received enough direct gradient supervision to act as a stable control signal, and $\mathbf{c} = \mathbf{t}$ suffices. When samples per item falls below this threshold, the target ID embedding is under-trained and we concatenate the masked sequence mean: $\mathbf{c} = [\mathbf{t};\, \bar{\mathbf{s}}]$. The architecture is unchanged across regimes; only the condition source switches. In our evaluation suite, the crossover falls between TaobaoAd (${\sim}59$ samples/item, where $\mathbf{c}{=}\mathbf{t}$ is selected and performs on par with item$+$sequence conditioning) and TAAC (${\sim}22$ samples/item, where $\mathbf{c}{=}[\mathbf{t};\bar{\mathbf{s}}]$ is optimal).

\subsection{Parameter and Compute Cost}
\label{sec:param_cost}

The combined QueryPos and LENS additions are negligible relative to the model's embedding tables: the total of ${\sim}90$K parameters amounts to less than 2\% of a typical embedding table budget (5--50M). The per-component breakdown is given in Appendix Table~\ref{tab:param_cost}. The compute overhead is similarly minimal: one matrix multiply of size $d_c \times qD$ for the gate, a small per-layer projection for TCPB, and $q \times L$ scalar additions to existing attention logits per layer. Neither module introduces new activation functions, normalisation layers, or sequential dependencies beyond what already exists in the cross-attention pathway.

\section{Experimental Setup}

\subsection{Datasets}

We evaluate on four datasets that span different density regimes, summarised in Table~\ref{tab:datasets}. The bolded \textbf{samples-per-item} row is the primary axis for the analysis. It spans nearly three orders of magnitude across the four datasets ($\sim\!1000 \to \sim\!1.1$) and serves as a proxy for target-ID embedding maturity. KuaiRec provides enough samples per item to learn a reliable target embedding, while KuaiRand does not; this contrast motivates the condition-source rule used by LENS.

\begin{table}[t]
\centering
\small
\setlength{\tabcolsep}{3pt}
\resizebox{\linewidth}{!}{%
\begin{tabular}{lrrrr}
\toprule
\textbf{Property} & \textbf{KuaiRec} & \textbf{TaobaoAd} & \textbf{TAAC} & \textbf{KuaiRand} \\
\midrule
Train samples & $\sim$10.0M & $\sim$21.9M & $\sim$153.2M & $\sim$5.06M \\
Eval samples & $\sim$2.5M & $\sim$3.1M & $\sim$21.1M & $\sim$43K \\
Item vocabulary & $\sim$10K & $\sim$375K & $\sim$7M & $\sim$4.4M \\
\textbf{Samples/item} & $\boldsymbol{\sim}$\textbf{1000} & $\boldsymbol{\sim}$\textbf{59} & $\boldsymbol{\sim}$\textbf{22} & $\boldsymbol{\sim}$\textbf{1.1} \\
Sequence length & 200 & 50 ($\times$2) & 100 & 200 \\
Non-seq features & 74 & 16 & 21 & 70 \\
Eval positive rate & $\sim$7.5\% & 5.01\% & 3.81\% & 17.41\% \\
\bottomrule
\end{tabular}
}%
\caption{Dataset characteristics, ordered by decreasing train samples per item (\textbf{bolded row}). KuaiRec~\citep{Gao2022KuaiRec} is the dense anchor ($\sim$1000 samples/item); TaobaoAd~\citep{AlibabaDisplayAd} is a medium-density transition zone; TAAC~\citep{Pan2026} is the sparse industrial anchor; KuaiRand~\citep{Gao2022KuaiRand} is an extremely sparse exposure log. The samples/item ratio spans roughly three orders of magnitude across the four datasets.}
\label{tab:datasets}
\end{table}

\textbf{KuaiRec}~\citep{Gao2022KuaiRec} has $\sim\!1000$ samples per item, so item embeddings are densely trained and target-history matching is reliable. We use positive-only histories capped at 200 items together with the full-side feature scope. The dataset is used as the dense reference for the robustness analysis because DIN's native sequence path is already strong here.

\textbf{TaobaoAd}~\citep{AlibabaDisplayAd} consists of 21.9M training samples with dual behaviour sequences and medium item density, and serves as a transition-zone calibration benchmark.

\textbf{TAAC}~\citep{Pan2026} is a sparse-vocabulary industrial dataset from the Tencent Advertising Algorithm Competition 2025, with $\sim\!22$ training samples per item. We use the time-split protocol with sequence length 100. In this regime item embeddings alone are unreliable, and non-sequential metadata becomes useful.

\textbf{KuaiRand}~\citep{Gao2022KuaiRand} is an exposure-log dataset with $\sim\!4.4$M items and $\sim\!1.1$ training samples per item. Its histories contain exposed-but-skipped items, clicks, strong engagement, and negative feedback. We adopt typed exposure histories with action-type embeddings as the default protocol; click-only filtering would discard many long-tail exposure tokens that are needed as sequence-side coverage for later target items. The standard-log train split and the random-log evaluation split have different positive rates, which reflects the exposure-log selection bias analysed in Section~\ref{sec:robustness}.

\subsection{Models and Controls}

We compare four model families: DIN~\citep{Zhou2018} as a target-aware attention baseline, MixFormer~\citep{MixFormer2026} and OneTrans~\citep{Chen2025} as unified sequential/non-sequential Transformer baselines, a selected HyFormer query-decoding reference, and the proposed QueryPos$+$LENS extensions applied to the HyFormer backbone. For each family, configurations are selected on seed 42 and then frozen for the three-seed confirmation; Appendix~\ref{app:baseline_configs} summarises the selected settings. HyFormer contains four optional switches that are ablated in Part I of Table~\ref{tab:unified_results}; we fix the Seq Pooling Tokens variant as the downstream \textbf{HyFormer} reference. All comparisons share the input protocol and use matched or selected capacity for each backbone. We also report two input scopes: the \emph{controlled} scope uses only item IDs and sequence features, while the \emph{full-side} scope adds all non-sequential metadata. The pair of scopes is used in Section~\ref{sec:robustness} to measure each model's sensitivity to non-sequential features.

For the cross-backbone portability study (Section~\ref{sec:portability}), we port both QueryPos and LENS to MixFormer and OneTrans, the two external baselines from Part~I. Neither backbone includes a dedicated query-specific position component, so QueryPos is the first such component they receive. We fix all backbone hyperparameters at the values used in Part~I and apply the same density-driven condition-source rule.

\subsection{Protocol}

All results use AUC as the primary metric. Main comparison tables report three-seed summaries (seeds 42, 123, 456). We use a two-phase protocol: Phase~1 explores configurations with seed 42; Phase~2 confirms selected comparisons over three seeds.

The LENS architecture is fixed as TCQG plus TCPB. The only variable is the condition source, which follows a pre-specified rule based on samples per item (Table~\ref{tab:datasets}): the target item embedding is used in dense and transition regimes, and an item$+$sequence condition (the target embedding concatenated with the masked sequence mean) is used in sparse regimes. Part III of Table~\ref{tab:unified_results} reports the condition-source ablation behind this rule.

\subsection{Implementation Details}

All models are implemented in PyTorch and trained on NVIDIA H20 GPUs with Adam optimizer (learning rate $10^{-3}$). Batch size is 4096 unless memory requires otherwise. Each baseline uses its own best-performing hyperparameters (layer count, hidden size, head count, etc.) rather than a shared configuration; Appendix~\ref{app:baseline_configs} lists the selected settings per backbone per dataset.

DIN uses the standard target-attention architecture with matched embedding dimension and a 3-layer MLP head; the ``full-side'' variant concatenates all non-sequential feature embeddings with the attention output before the MLP, creating the competing feature path analysed in Section~\ref{sec:robustness}. HyFormer uses 4 Transformer layers with 4 attention heads and a 3-layer MLP prediction head ($[256, 128, 1]$, ReLU); the Seq Pooling Tokens variant is selected as the reference (Section~\ref{sec:stage1_results}). MixFormer uses the selected capacity per dataset with target-in-dense enabled. OneTrans uses the bidirectional/no-causal-mask variant. On TaobaoAd, both MixFormer and OneTrans consume the primary behaviour sequence and target only; we treat these rows as conservative baselines.

\paragraph{KuaiRand typed-exposure protocol.} Action types are mapped to 4 categories (exposed/non-click, click/valid-play, strong engagement, hate) and embedded with a learnable action embedding table of dimension $D$. The action embedding is added to the item embedding before entering the sequence encoder: $\mathbf{e}_{\mathrm{token}} = \mathbf{e}_{\mathrm{item}} + \mathbf{e}_{\mathrm{action}}$.

\section{Results and Analysis}
\label{sec:results}

\newcommand{\auc}[2]{$#1{\color{gray}{\scriptscriptstyle\,\pm\,#2}}$}
\newcommand{\bauc}[2]{$\mathbf{#1}{\color{gray}{\scriptscriptstyle\,\pm\,#2}}$}
\newcommand{\sauc}[3]{$\mathbf{#1}^{\,\text{#3}}{\color{gray}{\scriptscriptstyle\,\pm\,#2}}$}

Table~\ref{tab:unified_results} presents the staged validation results across all four datasets, organised into five parts. Part~I validates Stage~1 by selecting the backbone reference. Part~II validates Stage~2 by establishing a position-aware reference. Parts~III--V validate Stage~3 by adding LENS target-conditioned query routing, selecting the condition source, and attributing the gains. We analyse each stage in turn below.

\begin{table*}[t]
\centering
\scriptsize
\setlength{\tabcolsep}{3.5pt}
\resizebox{\textwidth}{!}{%
\begin{tabular}{lrrrr}
\toprule
\textbf{Model / Ablation} & \textbf{KuaiRec} & \textbf{TaobaoAd} & \textbf{TAAC} & \textbf{KuaiRand} \\
\midrule
Samples/item (density proxy) & $\sim$1000 & $\sim$59 & $\sim$22 & $\sim$1.1 \\
\midrule
\multicolumn{5}{l}{\emph{Part I: external full-strength baselines and HyFormer switch ablation}} \\
\midrule
DIN$^{\dagger}$ & \auc{0.8281}{0.0039} & \auc{0.6383}{0.0002} & \auc{0.7480}{0.0003} & \auc{0.6511}{0.0006} \\
MixFormer$^{\dagger}$ & \auc{0.8341}{0.0046} & \auc{0.6363}{0.0004} & \auc{0.7527}{0.0018} & \auc{0.6629}{0.0052} \\
OneTrans$^{\dagger}$ & \auc{0.8333}{0.0013} & \auc{0.6405}{0.0012} & \auc{0.7544}{0.0046} & \auc{0.6622}{0.0038} \\
HyFormer (no switches) & \auc{0.8260}{0.0047} & \auc{0.6404}{0.0002} & \auc{0.7560}{0.0013} & \auc{0.6698}{0.0046} \\
\textbf{HyFormer + Seq Pooling Tokens} & \bauc{0.8328}{0.0037} & \bauc{0.6394}{0.0009} & \bauc{0.7553}{0.0021} & \bauc{0.6679}{0.0019} \\
HyFormer + NS Tokens in Query Boosting & \auc{0.8253}{0.0029} & \auc{0.6394}{0.0003} & \auc{0.7564}{0.0029} & \auc{0.6600}{0.0106} \\
HyFormer + Per-query FFN & \auc{0.8294}{0.0038} & \auc{0.6392}{0.0010} & \auc{0.7578}{0.0016} & \auc{0.6680}{0.0045} \\
HyFormer + Full Switches & \auc{0.8292}{0.0057} & \auc{0.6395}{0.0003} & \auc{0.7504}{0.0041} & \auc{0.6747}{0.0055} \\
\midrule
\multicolumn{5}{l}{\emph{Part II: static position mechanism on the selected HyFormer reference}} \\
\midrule
Global Pos Bias & \auc{0.8391}{0.0032} & \auc{0.6392}{0.0010} & \auc{0.7571}{0.0017} & \auc{0.6892}{0.0055} \\
Abs Pos Emb & \auc{0.8397}{0.0033} & \auc{0.6394}{0.0011} & \auc{0.7591}{0.0011} & \auc{0.6658}{0.0039} \\
\textbf{Query-Specific Pos Bias} & \bauc{0.8395}{0.0030} & \bauc{0.6392}{0.0009} & \bauc{0.7570}{0.0024} & \bauc{0.6900}{0.0018} \\
\midrule
\multicolumn{5}{l}{\emph{Part III: choosing the target condition for LENS}} \\
\midrule
LENS w/ Item Condition & \auc{0.8418}{0.0012} & \auc{0.6403}{0.0007} & \auc{0.7569}{0.0021} & \auc{0.6900}{0.0035} \\
LENS w/ Item+Seq Condition & \auc{0.8401}{0.0032} & \auc{0.6403}{0.0012} & \auc{0.7589}{0.0019} & \auc{0.6945}{0.0027} \\
\textbf{LENS w/ Selected Condition} & \sauc{0.8418}{0.0012}{i} & \sauc{0.6403}{0.0007}{i} & \sauc{0.7589}{0.0019}{i+s} & \sauc{0.6945}{0.0027}{i+s} \\
\midrule
\multicolumn{5}{l}{\emph{Part IV: attributing the two target-conditioned modules}} \\
\midrule
LENS w/o TCPB & \auc{0.8388}{0.0031} & \auc{0.6395}{0.0004} & \auc{0.7575}{0.0004} & \auc{0.6948}{0.0035} \\
LENS w/o TCQG & \auc{0.8383}{0.0048} & \auc{0.6409}{0.0010} & \auc{0.7565}{0.0021} & \auc{0.6952}{0.0015} \\
Full LENS & \auc{0.8418}{0.0012} & \auc{0.6403}{0.0007} & \auc{0.7589}{0.0019} & \auc{0.6945}{0.0027} \\
\midrule
\multicolumn{5}{l}{\emph{Part V: LENS module gain and total LENS gain}} \\
\midrule
$\Delta$ modules over QueryPos base & $\mathbf{+0.0023}$ & $\mathbf{+0.0011}$ & $\mathbf{+0.0019}$ & $\mathbf{+0.0045}$ \\
$\Delta$ final LENS vs HyFormer ref. & $\mathbf{+0.0090}$ & $\mathbf{+0.0009}$ & $\mathbf{+0.0036}$ & $\mathbf{+0.0266}$ \\
$\Delta$ final LENS vs DIN$^{\ddagger}$ & $\mathbf{+0.0137}$ & $\mathbf{+0.0020}$ & $\mathbf{+0.0109}$ & $\mathbf{+0.0434}$ \\
\bottomrule
\end{tabular}
}%
\caption{Unified main results and ablations (HyFormer backbone). Each result row reports AUC mean with a smaller gray three-seed std. Bold rows mark the selected path: HyFormer + Seq Pooling Tokens, Query-Specific Pos Bias, and LENS w/ Selected Condition. \textsuperscript{$\dagger$}~Faithful re-implementation: DIN~\citep{Zhou2018}, MixFormer~\citep{MixFormer2026}, OneTrans~\citep{Chen2025}. \textsuperscript{$\ddagger$}~DIN uses the full-side scope; on KuaiRec, Section~\ref{sec:robustness} shows non-sequential features hurt DIN by $-0.010$ AUC. In Part III, the superscript on each Selected Condition cell indicates the chosen source: \textsuperscript{i}~item-only, \textsuperscript{i+s}~item+sequence.}
\label{tab:unified_results}
\end{table*}

\subsection{Stage 1: Backbone Reference Selection}
\label{sec:stage1_results}

Part~I of Table~\ref{tab:unified_results} shows that no single HyFormer switch configuration dominates all four datasets. The absolute spread among variants is modest (at most 0.0075 AUC on KuaiRec). We select \textbf{HyFormer + Seq Pooling Tokens} as the backbone reference: its mean gap to the per-dataset optimum is only 0.0026 AUC, with the lowest worst-case deficit and the lowest cross-dataset variance (std $\leq 0.0037$ on all benchmarks).

\begin{figure}[t]
\centering
\includegraphics[width=\linewidth]{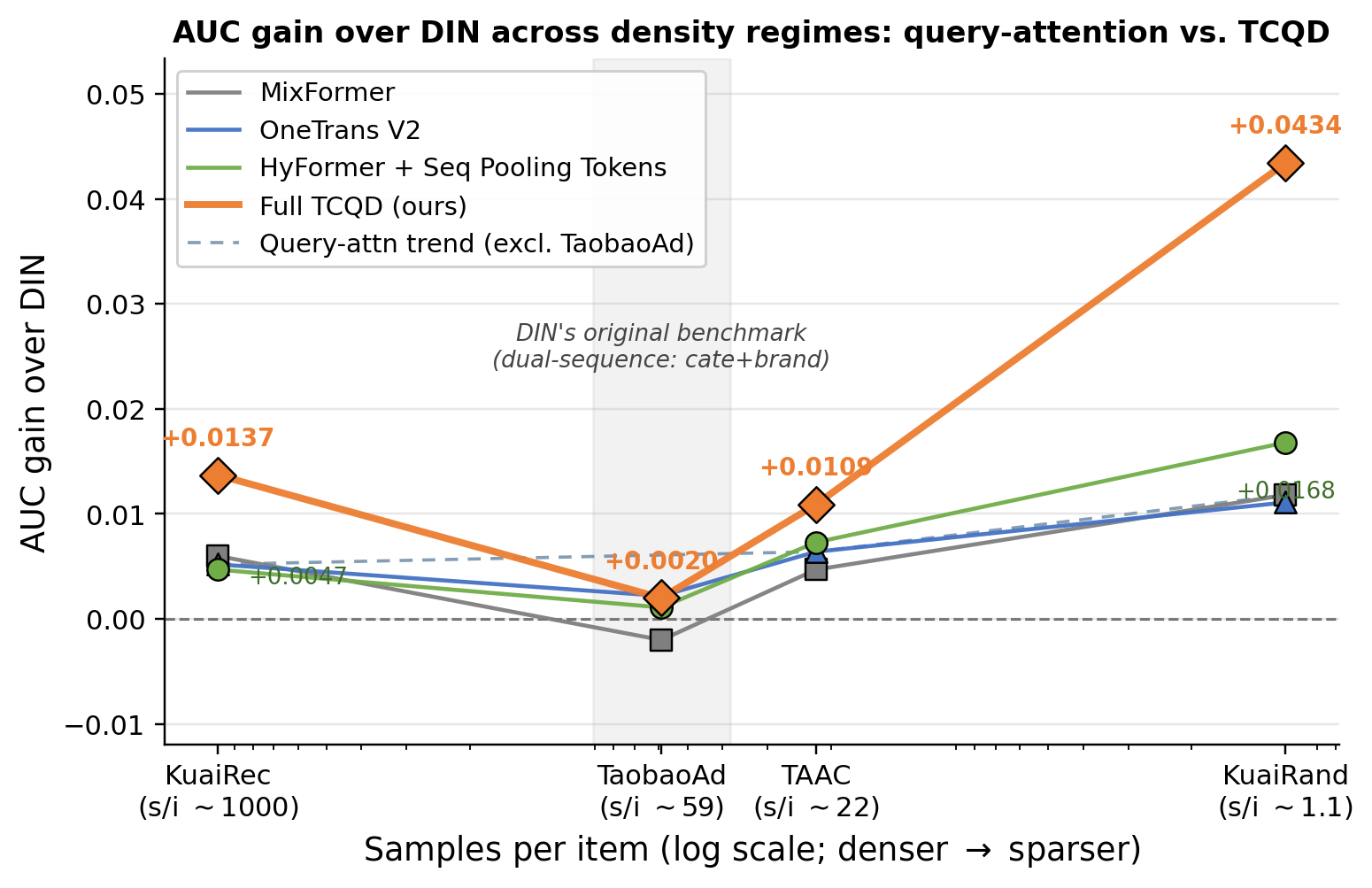}
\caption{AUC gain over DIN as a function of item density. Latent-query models show increasing advantage as density decreases from KuaiRec ($\sim$1000) to KuaiRand ($\sim$1.1).}
\label{fig:query_attention_gain}
\end{figure}

Figure~\ref{fig:query_attention_gain} shows that all latent-query models exhibit increasing advantage over DIN as item density decreases. On dense KuaiRec, DIN remains competitive; the benefit of richer cross-attention emerges in sparser regimes where item embeddings are unreliable.

\subsection{Stage 2: Position Mechanism Validation}
\label{sec:stage2_results}

Part~II compares three static position mechanisms on the backbone reference. All three provide substantial gains over the unpositioned backbone (compare to Part~I), confirming that explicit position structure is a first-order contributor in latent-query architectures. Among the three, no mechanism uniformly dominates. Abs Pos Emb achieves the highest AUC on KuaiRec and TAAC but drops substantially on KuaiRand (std 0.0039). Global Pos Bias is competitive on KuaiRand but has higher variance (0.0055). \textbf{Query-Specific Pos Bias} (QueryPos) trails the per-dataset optimum by at most 0.0021 AUC on every benchmark while maintaining the lowest cross-dataset variance. We adopt QueryPos as the position-aware reference.

\subsection{Stage 3: Target-Conditioned Query Routing}
\label{sec:stage3_results}

Parts~III--V of Table~\ref{tab:unified_results} evaluate LENS target-conditioned query routing atop the QueryPos reference.

\paragraph{Part III: Condition source selection.} The condition source determines what information is injected into the position bias and query gate. Part~III compares two sources---item-only (the target item embedding) and item+sequence (the target embedding concatenated with the masked sequence mean)---and reports the density-driven selection rule. On KuaiRec ($\sim$1000 samples/item), the item-only condition exceeds item+sequence; on TaobaoAd ($\sim$59), the two are tied and item-only is selected under the rule, yielding the superscript \textsuperscript{i}. On TAAC ($\sim$22) and KuaiRand ($\sim$1.1), item+sequence provides the gain, yielding \textsuperscript{i+s}. The transition boundary therefore falls near $\sim$50 samples/item, which aligns with the intuition that reliable item embeddings remove the need for additional sequence context in the conditioning signal.

\begin{figure}[t]
\centering
\includegraphics[width=\linewidth]{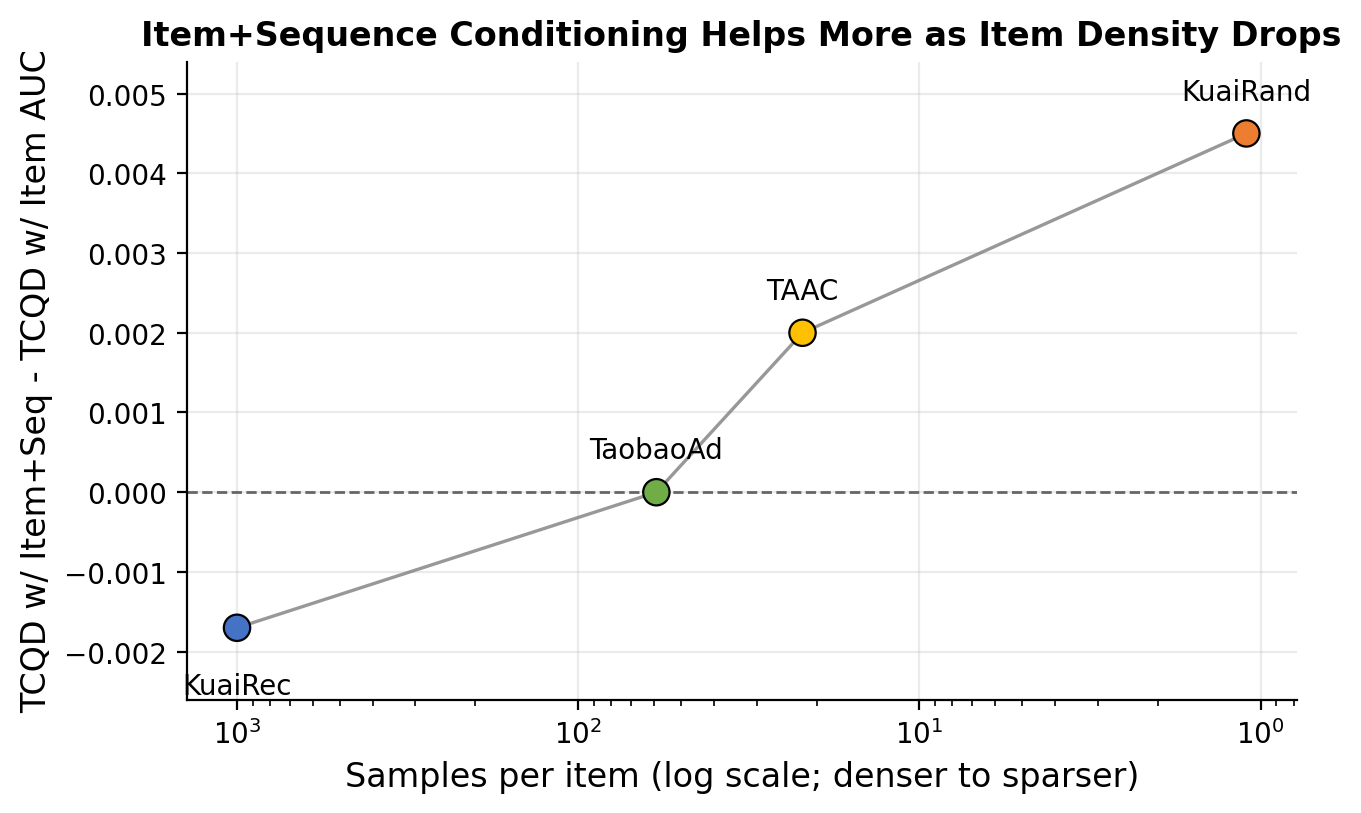}
\caption{Condition source AUC gain (item+seq minus item-only) as a function of item density. The crossover at approximately 50 samples/item separates the dense item-only regime from the sparse item+sequence regime. TAAC and KuaiRand benefit from the richer condition; KuaiRec and TaobaoAd do not.}
\label{fig:condition_source_density}
\end{figure}

Figure~\ref{fig:condition_source_density} plots the condition-source AUC gain (item+sequence minus item-only) against item density. The crossover near $\sim$50 samples/item separates datasets where the enriched condition helps (TAAC, KuaiRand) from those where it is neutral or slightly harmful (KuaiRec, TaobaoAd), with TaobaoAd lying near the boundary. This empirical boundary motivates the pre-specified rule used throughout the paper: apply item-only conditioning when samples/item exceeds $\sim$50, and item+sequence conditioning otherwise.

\paragraph{Part IV: Module attribution.} Part~IV decomposes the full LENS system into its two components---Target-Conditioned Position Bias (TCPB) and Target-Conditioned Query Gate (TCQG)---by removing each in turn. On TaobaoAd and KuaiRand, the system without TCQG (i.e., TCPB alone) slightly exceeds the full model: 0.6409 and 0.6952 respectively, versus 0.6403 and 0.6945 for Full LENS. This suggests that TCPB captures most of the Stage~3 gain in medium-to-sparse density regimes, where position-dependent target modulation provides the primary benefit and the additional gate contributes less. Conversely, on KuaiRec and TAAC, removing either module degrades performance relative to Full LENS (0.8388 and 0.7575 without TCPB; 0.8383 and 0.7565 without TCQG), indicating both modules contribute. The complementary pattern confirms that TCPB and TCQG address different aspects of the target-conditioned computation: TCPB modulates \emph{where} each query retrieves history, while TCQG modulates \emph{which} queries are active for a given candidate.

\paragraph{Part V: Delta summary.} The three delta rows quantify cumulative gains at different reference points. The module delta ($\Delta$ modules over QueryPos base) isolates the value of LENS modules alone, ranging from +0.0011 on TaobaoAd to +0.0045 on KuaiRand. The total LENS delta versus the HyFormer reference (without position or conditioning) captures the joint QueryPos$+$LENS contribution: +0.0090 on KuaiRec, +0.0009 on TaobaoAd, +0.0036 on TAAC, and +0.0266 on KuaiRand. Finally, the delta versus DIN demonstrates the end-to-end advantage of the full QueryPos$+$LENS system over a conventional target-attention baseline, reaching +0.0434 AUC on KuaiRand---the sparsest regime where latent-query modelling with target conditioning yields the largest benefit.

\begin{figure*}[t]
\centering
\includegraphics[width=0.85\textwidth]{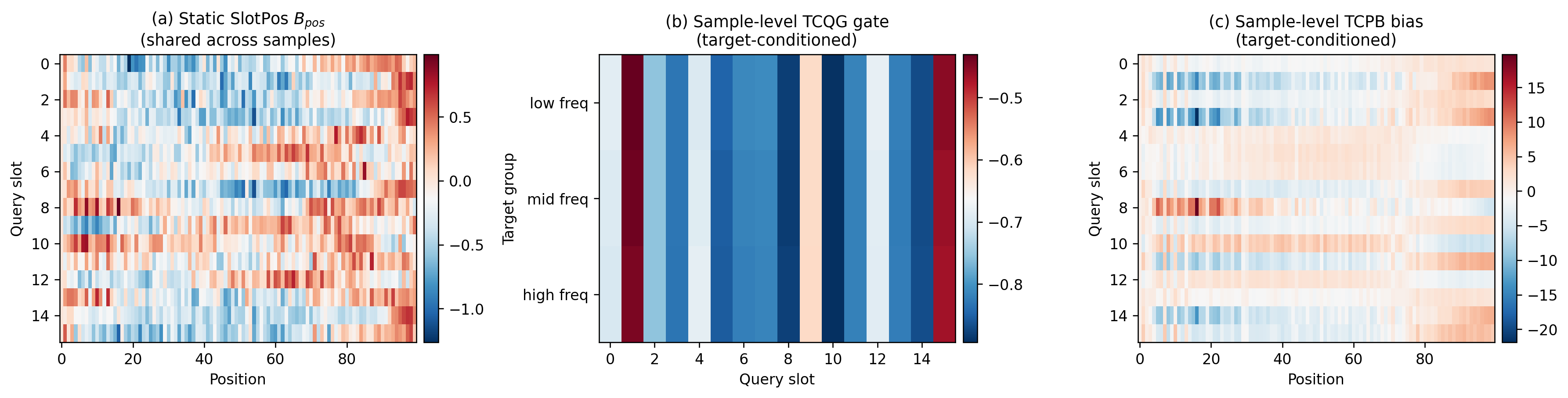}
\caption{Attention heatmaps on TAAC. \textbf{(a)} Static QueryPos bias shared across samples. \textbf{(b)} TCQG gate aggregated by target-item frequency tertile. \textbf{(c)} Sample-averaged TCPB bias. The static map is target-independent; the conditioned maps show per-candidate redistribution of attention mass.}
\label{fig:taac_heatmaps}
\end{figure*}

Figure~\ref{fig:taac_heatmaps} provides qualitative evidence for the conditioning mechanism. On TAAC, the static QueryPos bias produces a position-dependent but target-independent attention pattern. After target conditioning, the heatmap shows clear redistribution of attention mass: for one target category the model concentrates on recent positions (80--100), while for a different target the attention spreads across mid-sequence positions (40--70). This qualitative shift confirms that LENS enables the model to adaptively route attention in a target-specific manner, which the static position bias alone cannot achieve.

\subsection{Cross-Backbone Portability}
\label{sec:portability}

The portability results are the primary evidence that the proposed additions transfer within the latent-query family. If the same Stage~2 and Stage~3 additions improve multiple latent-query backbones, then the design is better understood as a portable recipe than as a single-backbone variant. The QueryPos+LENS design is therefore tested beyond HyFormer. Both components require only that the backbone exposes a query/key or cross-attention pathway where a position bias and a target-conditioned routing module can be injected. We port both QueryPos and LENS to MixFormer~\citep{MixFormer2026} and OneTrans~\citep{Chen2025}, the two external full-strength baselines from Part~I. Neither backbone includes a dedicated query-specific position component: MixFormer uses raw content attention in its cross-attention blocks, and OneTrans relies solely on self-attention over a concatenated token stream without position embeddings.

\begin{table*}[t]
\centering
\small
\setlength{\tabcolsep}{4pt}
\begin{tabular}{llrrrrrr}
\toprule
\textbf{Backbone} & \textbf{Dataset} & \textbf{Baseline} & \textbf{+QueryPos} & \textbf{+QueryPos+LENS} & $\boldsymbol{\Delta}$ \textbf{QueryPos} & $\boldsymbol{\Delta}$ \textbf{LENS} & $\boldsymbol{\Delta}$ \textbf{Total} \\
                  &                  & (3-seed)          & (3-seed)           & (3-seed)               &        &           & vs base \\
\midrule
MixFormer    & KuaiRec  & \auc{0.8341}{0.0046} & \auc{0.8391}{0.0053} & \bauc{0.8399}{0.0045} & $+0.0050$ & $+0.0008$ & $\mathbf{+0.0058}$ \\
MixFormer    & TaobaoAd & \auc{0.6363}{0.0004} & \auc{0.6368}{0.0002} & \bauc{0.6377}{0.0006} & $+0.0005$ & $+0.0009$ & $\mathbf{+0.0014}$ \\
MixFormer    & TAAC     & \auc{0.7527}{0.0018} & \auc{0.7524}{0.0030} & \bauc{0.7534}{0.0023} & $-0.0003$ & $+0.0010$ & $\mathbf{+0.0007}$ \\
MixFormer    & KuaiRand & \auc{0.6629}{0.0052} & \auc{0.6636}{0.0053} & \bauc{0.6683}{0.0039} & $+0.0007$ & $+0.0047$ & $\mathbf{+0.0054}$ \\
\midrule
OneTrans  & KuaiRec  & \auc{0.8333}{0.0013} & \auc{0.8403}{0.0020} & \bauc{0.8472}{0.0047} & $+0.0070$ & $+0.0069$ & $\mathbf{+0.0139}$ \\
OneTrans  & TaobaoAd & \auc{0.6405}{0.0012} & \auc{0.6416}{0.0011} & \bauc{0.6439}{0.0003} & $+0.0011$ & $+0.0023$ & $\mathbf{+0.0034}$ \\
OneTrans  & TAAC     & \auc{0.7544}{0.0046} & \auc{0.7542}{0.0055} & \bauc{0.7556}{0.0013} & $-0.0002$ & $+0.0014$ & $\mathbf{+0.0012}$ \\
OneTrans  & KuaiRand & \auc{0.6622}{0.0038} & \auc{0.6636}{0.0025} & \bauc{0.6687}{0.0025} & $+0.0014$ & $+0.0052$ & $\mathbf{+0.0065}$ \\
\bottomrule
\end{tabular}
\caption{Cross-backbone portability. Baselines are reproduced from Part~I. All values are 3-seed mean$\,\pm\,$std. \textbf{Bold} entries indicate positive $\Delta$~Total. $\Delta$~QueryPos and $\Delta$~LENS decompose the total into static position prior and target-conditioned residual contributions.}
\label{tab:portability}
\end{table*}

Table~\ref{tab:portability} reports the per-component decomposition across all eight backbone$\times$dataset cells. All eight cells achieve positive total-gain point estimates ($\Delta$~Total). The largest gain appears on OneTrans$\times$KuaiRec ($+0.0139$), where QueryPos and LENS contribute nearly equally ($+0.0070$ and $+0.0069$). This balanced decomposition reflects OneTrans's lack of native position encoding: both the static position prior and target-conditioned modulation address genuine architectural gaps.

On TAAC, a consistent pattern appears across both backbones: QueryPos contributes near-zero deltas ($-0.0003$ for MixFormer, $-0.0002$ for OneTrans), while LENS accounts for the positive total-gain point estimates ($+0.0010$ and $+0.0014$). This is consistent with TAAC's shorter history truncation ($L{=}100$): when sequences are shorter, explicit position structure adds less marginal information, but target conditioning remains valuable for routing attention based on the candidate.

MixFormer shows a similar QueryPos-weak, LENS-strong pattern on KuaiRand and TAAC. This reflects MixFormer's per-head dense architecture, where the target embedding is already integrated into every attention head through the initial projection~\citep{MixFormer2026}. The pre-existing target-content coupling limits the marginal value of an additive position prior, but does not subsume the benefit of explicit target-conditioned modulation. On KuaiRec, where item embeddings are well-trained, QueryPos does contribute meaningfully ($+0.0050$), consistent with position structure being more informative when content embeddings are reliable.

The practical implication is that QueryPos and LENS function as a modular design recipe: apply QueryPos as a low-risk position baseline, then optionally layer LENS with the density-driven condition-source rule. Both are additive in parameter cost ($<$2\% of embedding tables) and preserve the pre-existing attention pattern at training start through zero-initialised residual connections.

Taken together, Table~\ref{tab:portability} and the HyFormer staged analysis support the same conclusion: interaction granularity is not exhausted by selecting a single reference backbone. The positive total-gain point estimates in all twelve backbone$\times$dataset cells show that the Stage~2 and Stage~3 additions transfer across multiple latent-query backbones, which is why we view portability as primary evidence that the design is not tied to one architecture.

\section{Supporting Diagnostics}

The preceding sections present the main LENS architecture result. This section collects two diagnostics that are relevant for deploying sequential CTR models but do not themselves belong to the LENS architecture. The first concerns non-sequential features in DIN-like models on dense item vocabularies, where they can create a shortcut that competes with the sequence-attention path. The second concerns history protocols on sparse-vocabulary logs such as KuaiRand, where positive-only filtering removes the exposure tokens needed for embedding coverage.

\subsection{More Features Are Not Always Better for DIN-like Models}
\label{sec:robustness}

\begin{figure}[t]
  \centering
  \includegraphics[width=\linewidth]{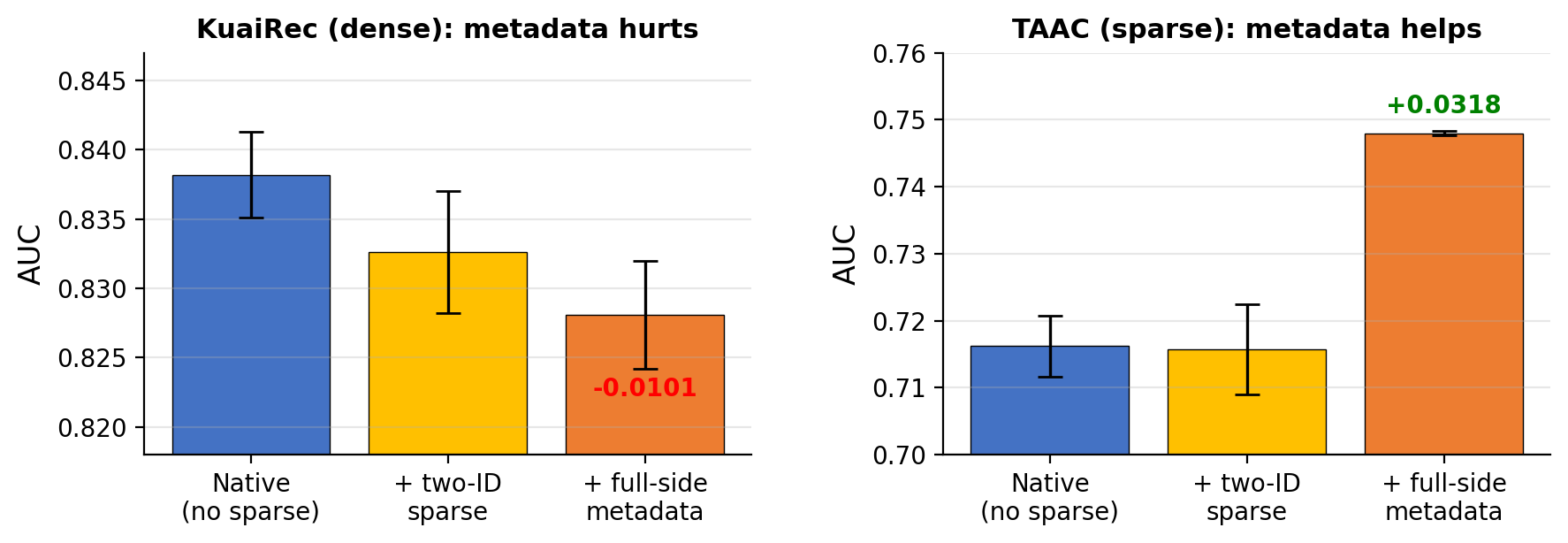}
  \vspace{-1mm}
  \scriptsize
  \resizebox{\linewidth}{!}{%
  \begin{tabular}{lrr}
  \toprule
  \textbf{DIN Feature Path} & \textbf{KuaiRec} & \textbf{TAAC} \\
  \midrule
  Native (no sparse) & $\mathbf{0.8382 \pm 0.0031}$ & $0.7162 \pm 0.0046$ \\
  + two-ID sparse & $0.8326 \pm 0.0044$ & $0.7157 \pm 0.0067$ \\
  + full-side metadata & $0.8281 \pm 0.0039$ & $\mathbf{0.7480 \pm 0.0003}$ \\
  \bottomrule
  \end{tabular}
  }%
  \caption{DIN feature-path analysis (3-seed). On dense KuaiRec, adding extra non-sequential information hurts DIN: two ID features already reduce AUC, and full-side metadata drops further. On sparse TAAC, the same full-side metadata is strongly helpful.}
  \label{fig:din_feature_path}
\end{figure}

Figure~\ref{fig:din_feature_path} gives a simple diagnostic on whether additional feature paths help DIN-like target-attention models. On KuaiRec, DIN's native target--history path is already strong because the item vocabulary is dense ($\sim$1000 samples per item) and the model can learn reliable item-to-history matching from the sequence alone. Adding the two-ID sparse path lowers AUC from $0.8382$ to $0.8326$, and adding full-side metadata lowers it further to $0.8281$. The extra features do not act purely as information; they introduce an MLP path that can memorise coarse user--item co-occurrence and weakens the pressure to use the sequence-attention path.

TAAC reverses the situation. With $\sim$22 samples per item, DIN's native path is much weaker: native DIN reaches $0.7162$ and the two-ID sparse path is statistically tied. Full-side metadata raises AUC to $0.7480$ with low variance. Metadata is therefore informative when item embeddings are sparse and shared side fields generalise across items, and harmful when item embeddings are already mature and the sequence path is reliable.

\begin{table}[t]
\centering
\small
\resizebox{\linewidth}{!}{%
\begin{tabular}{lrrr}
\toprule
\textbf{Model} & \textbf{Controlled (no non-seq)} & \textbf{Full-side} & \textbf{$\Delta$} \\
\midrule
DIN & $0.8382{\color{gray}{\scriptscriptstyle\,\pm\,0.0031}}$ & $0.8281{\color{gray}{\scriptscriptstyle\,\pm\,0.0039}}$ & $-0.0101$ \\
HyFormer & $0.8338{\color{gray}{\scriptscriptstyle\,\pm\,0.0025}}$ & $0.8328{\color{gray}{\scriptscriptstyle\,\pm\,0.0037}}$ & $-0.0010$ \\
LENS (w/o TCPB) & $0.8388{\color{gray}{\scriptscriptstyle\,\pm\,0.0014}}$ & $0.8388{\color{gray}{\scriptscriptstyle\,\pm\,0.0031}}$ & $\mathbf{0.0000}$ \\
\bottomrule
\end{tabular}
}%
  \caption{KuaiRec feature-path diagnostic using LENS without Target-Conditioned Position Bias (TCPB), with 3-seed means and smaller gray std: controlled (sequence + item IDs only) vs full-side (all non-sequential metadata enabled). DIN drops $-0.010$ when non-sequential features are added; HyFormer drops $-0.001$; the gate-only LENS variant does not degrade in this diagnostic setting.}
\label{tab:robustness}
\end{table}

Table~\ref{tab:robustness} asks whether query-decoding models share the full-side shortcut behaviour on KuaiRec. DIN drops by $-0.0101$ AUC from the controlled to the full-side input, HyFormer drops by $-0.0010$, and LENS without TCPB reaches the same rounded mean AUC in both settings. We isolate TCQG in this probe because Figure~\ref{fig:din_feature_path} diagnoses a feature-path shortcut: the question is whether target-conditioned query filtering prevents non-sequential metadata from overwhelming the sequence path. TCPB is a position-retrieval module and would add a second, position-level conditioning effect orthogonal to the diagnostic.

Among the three rows, the gate-only LENS variant attains the highest AUC in both settings ($0.8388$ controlled, $0.8388$ full-side), consistent with TCQG acting as a relevance filter on the metadata path. The corresponding numbers for Full LENS are reported in Appendix~\ref{app:full_lens_robustness}; they are not used as the main robustness claim because, on dense KuaiRec, TCPB strengthens the mature target-ID conditioning path and increases sensitivity to the controlled item-ID-only setting, conflating the two effects.

HyFormer also degrades, although less than DIN. Its non-sequential features enter through QueryGen and mix into the query initialisation; without a target-specific filter, the sequence-derived query representation is diluted by metadata that may be irrelevant to the current candidate. The mean-pooled sequence injection in QueryGen reduces but does not remove this dilution, since it cannot suppress individual metadata dimensions per candidate. TCQG addresses the dilution directly: it conditions the query initialisation on the target embedding before any sequence aggregation, leaving metadata dimensions aligned with the candidate amplified and unrelated dimensions damped toward the initialisation value. This is consistent with the ordering in Table~\ref{tab:robustness}: DIN has no filter ($-0.0101$), HyFormer has partial mitigation through sequence injection ($-0.0010$), and TCQG provides explicit target-conditioned filtering ($0.0000$).

\subsection{For Long-Tail Items, Do Not Use Positive-Only Histories}
\label{sec:action_results}

\begin{figure}[t]
  \centering
  \includegraphics[width=\linewidth]{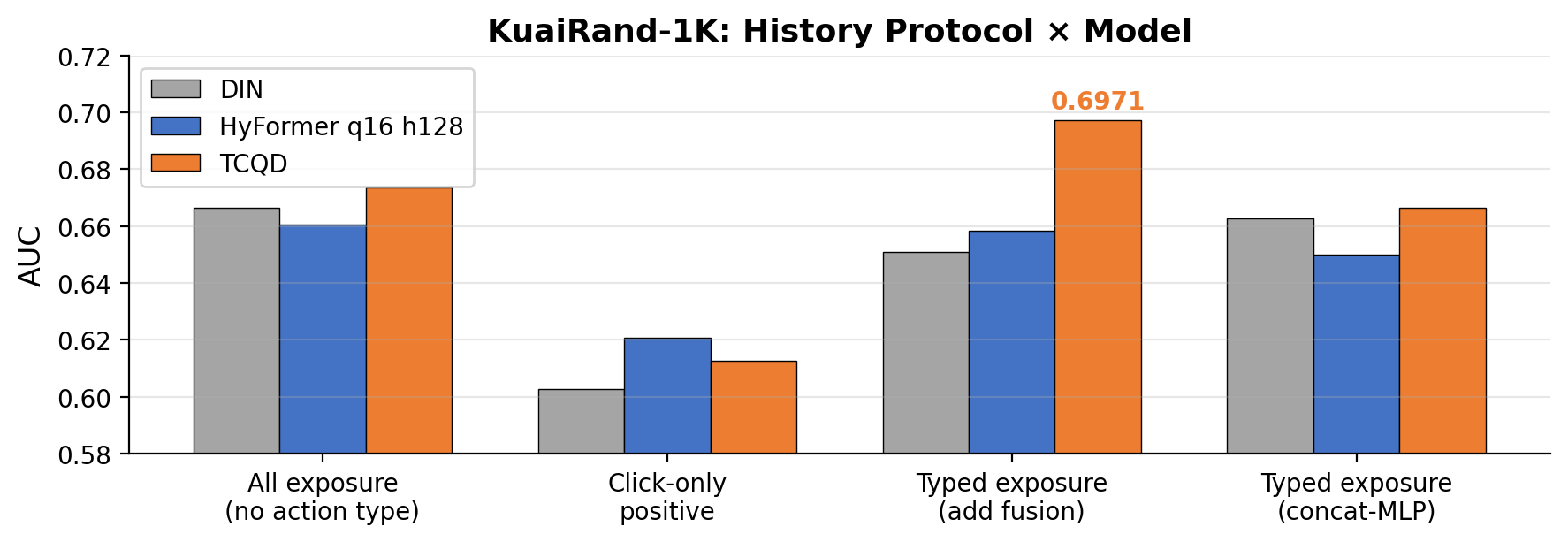}
  \vspace{-1mm}
  \scriptsize
  \resizebox{\linewidth}{!}{%
  \begin{tabular}{llrrr}
  \toprule
  \textbf{Protocol} & \textbf{Fusion} & \textbf{DIN} & \textbf{HyFormer} & \textbf{LENS} \\
  \midrule
  All exposure & none & 0.6664 & 0.6606 & 0.6736 \\
  Click-only & none & 0.6026 & 0.6207 & 0.6126 \\
  Typed exposure & add & 0.6509 & 0.6583 & $\mathbf{0.6971}$ \\
  Typed exposure & concat & 0.6628 & 0.6501 & 0.6664 \\
  \bottomrule
  \end{tabular}
  }%
  \caption{KuaiRand history-protocol comparison (seed 42; the selected typed-add LENS row is confirmed over 3 seeds in Table~\ref{tab:unified_results}). Click-only histories lose embedding coverage for long-tail target items; typed exposure histories preserve coverage while adding behavioural disambiguation.}
  \label{fig:action_semantics}
\end{figure}

Figure~\ref{fig:action_semantics} compares four KuaiRand history protocols. Click-only histories are substantially worse than all-exposure histories: LENS falls from $0.6736$ AUC with all exposures to $0.6126$ with click-only, and DIN and HyFormer show the same direction. The drop is large enough to be consistent with a protocol-level coverage problem rather than a modelling fluctuation.

A plausible dominant factor is embedding coverage. KuaiRand has $\sim$4.4M items and $\sim$1.1 training samples per item. Restricting the history to clicked positives removes most long-tail items from the sequence side during training, leaving their embeddings without sequence-side gradient. Those items still appear as negative targets in evaluation, but the model cannot score them against latent queries because the corresponding key embeddings are essentially untrained. All-exposure histories recover much of the lost AUC by retaining displayed-but-skipped items as sequence tokens.

All-exposure histories conflate behaviour, however: skipped and clicked items carry identical representations, so the sequence cannot expose interest signals. Typed exposure histories combine the two requirements. They keep all exposure tokens, so embedding coverage is preserved, and they add a lightweight action-type embedding so that exposed-but-skipped tokens are distinguishable from positively engaged ones. Under this protocol LENS reaches $0.6971$, well above HyFormer ($0.6583$) and DIN ($0.6509$). The gap over HyFormer ($+0.0388$) is the largest among our four datasets, consistent with target-conditioned query and position modulation being most useful when histories contain heterogeneous behaviours.

The fusion comparison is also informative. LENS prefers additive fusion ($e_{\mathrm{item}}+e_{\mathrm{action}}$), while DIN benefits from concat-MLP fusion. A plausible explanation is that query decoders rely on a stable token space in which action type modulates the item embedding smoothly, whereas DIN's item-level attention can absorb a richer per-token representation; the concat-MLP projection produces a token space that the query decoder does not fully recover.

\section{Discussion}

\paragraph{Evidence for separable design layers.}
A useful way to interpret the experimental results is to view target--history interaction in latent-query sequential CTR as a stack of design layers rather than as a single architecture choice. In our experiments, the backbone bottleneck explains the largest variance, QueryPos adds a static position prior orthogonal to that bottleneck, and LENS adds candidate-specific modulation on top. The gains from Stage~2 and Stage~3 are approximately additive across the portability cells, which is why we interpret the results as evidence for a portable design recipe rather than for a single architecture-specific trick.

\paragraph{QueryPos as a portable position prior.}
The latent-query CTR backbones evaluated here do not explicitly model query-specific position preferences. Adding a per-query learnable position bias yields consistent improvements, especially on datasets with longer sequences. QueryPos adds negligible parameters ($N_L \cdot q \cdot L_{\max}$, typically $<$10K) and requires only a single additive term in the attention logits, making it a natural Stage~2 baseline before any target conditioning is introduced.

\paragraph{Stage~3 gains are portable even when their balance shifts.}
Across both MixFormer and OneTrans, the Stage~3 gain from LENS remains positive in every reported cell, and on MixFormer it is often larger than the Stage~2 gain from QueryPos. MixFormer's per-head projection already incorporates the target into every dense head~\citep{MixFormer2026}, limiting the marginal value of an additional query gate. None of the evaluated backbones natively supports target-conditioned position retrieval, so the Stage~3 residual still fills a gap within the tested latent-query family. This helps explain why the target-conditioned gain remains portable even when the backbone's native target pathway changes.

\paragraph{Density-driven condition source.}
The crossover near ${\sim}50$ samples per item separates item-only from item$+$sequence conditioning. Above this threshold the target-ID embedding is reliable; below it the masked sequence mean supplements the undertrained embedding. We do not claim that this boundary is universal, but our experiments suggest that samples-per-item is a useful one-dimensional proxy for deciding how much sequence context should enter the conditioning signal.

\paragraph{Implications for future architecture design.}
The staged view is useful beyond the particular LENS instantiation studied here, especially for latent-query CTR backbones. Rather than treating interaction granularity as a fixed architectural side effect, these models can choose a bottleneck, add a position-aware reference, and then evaluate target-conditioned modulation as a separate layer. More broadly, the staged analysis offers a way to evaluate new components in isolation instead of attributing all gains to a monolithic architecture revision.

\paragraph{Practical guidelines.}
A practical deployment recipe follows directly from the stages. First, tune backbone capacity switches and select the interaction bottleneck. Second, add QueryPos as a low-risk position prior; it helps most on longer sequences ($L \geq 200$) and is often neutral rather than harmful on shorter ones. Third, optionally layer LENS with zero initialisation and the density-driven condition-source rule. On MixFormer-like architectures, TCPB alone may already capture most of the Stage~3 gain. We nevertheless treat the full LENS system as the default portable recipe: TCQG remains useful on dense regimes and underlies the target-conditioned filtering effect highlighted by the feature-path diagnostic in Section~\ref{sec:robustness}. For sparse-vocabulary exposure logs, retain all exposure tokens with action-type embeddings rather than filtering to positives.

\section{Conclusion}

We showed that interaction granularity---the level at which the candidate modulates history aggregation---is a designable, separable dimension within latent-query sequential CTR backbones rather than a fixed consequence of backbone choice. Within this staged view, QueryPos provides a portable position-aware reference and LENS provides target-conditioned residuals that restore candidate-specific control through query activation and position retrieval. Across HyFormer, MixFormer, and OneTrans on four datasets spanning ${\sim}10^3$ to ${\sim}1.1$ samples per item, all twelve backbone$\times$dataset cells show positive total-gain point estimates, supporting portability within the latent-query family as primary evidence that the design is not tied to one architecture. We further identify a density-driven condition-source rule---item-only above ${\sim}50$ samples per item, item$+$sequence below---and two practical diagnostics on feature-path shortcuts and typed-exposure histories. Each added component costs $<$2\% of embedding tables, making the approach practical for incremental adoption.

\bibliographystyle{ACM-Reference-Format}

\clearpage
\appendix
\section{Extended Results and Configurations}
\label{app:full_grids}

This appendix collects baseline configurations, parameter cost details, supporting visualisations, and per-seed details for the DIN feature-path diagnostic.

\begin{table*}[t]
\centering
\small
\begin{tabular}{lrr}
\toprule
\textbf{Component} & \textbf{Parameters} & \textbf{Example} \\
\midrule
Query-Specific Position Bias ($B_{\mathrm{pos}}$) & $N_L \cdot q \cdot L_{\max}$ & $4 \times 12 \times 200 = 9.6$K \\
Target-Conditioned Query Gate & $q D^2$ & $12 \times 64^2 = 49$K \\
Target-Conditioned Position Bias & $N_L(qDr + L_{\max}r)$ & $4(12 \times 64 \times 8 + 200 \times 8) \approx 31$K \\
\midrule
Total LENS & & $\sim$90K \\
Full embedding tables & & $\sim$5--50M \\
\bottomrule
\end{tabular}
\caption{Parameter cost for a typical configuration ($q{=}12$, $D{=}64$, $L_{\max}{=}200$, $N_L{=}4$, $r{=}8$, item-only condition $d_c{=}D$). The combined QueryPos and LENS additions constitute less than 2\% of embedding table parameters (typically 5--50M).}
\label{tab:param_cost}
\end{table*}

\subsection{Baseline Configuration Summary}
\label{app:baseline_configs}

Table~\ref{tab:baseline_configs} summarises the per-dataset configurations used for the four backbones in the main comparison. The settings are selected during seed 42 exploration under the current protocol and then fixed before three-seed confirmation. ``No causal mask'' denotes the bidirectional OneTrans variant, selected because the causal-mask setting was worse in offline CTR evaluation. TaobaoAd MixFormer and OneTrans currently use the primary behaviour sequence and target only. Complete configuration files and training records will be released as supplementary artifacts.

\begin{table}[t]
\centering
\small
\setlength{\tabcolsep}{4pt}
\begin{tabularx}{\linewidth}{@{}llX@{}}
\toprule
\textbf{Model} & \textbf{Dataset} & \textbf{Configuration} \\
\midrule
DIN & KuaiRec & Full-side target attention, maxlen 200 \\
    & TaobaoAd & Dual-seq full-side DIN, maxlen $50{\times}2$ \\
    & TAAC & Full-side DIN, maxlen 100 \\
    & KuaiRand & Full-side DIN with additive action-type embedding, maxlen 200 \\
\midrule
HyFormer & KuaiRec & q12 h64, Seq Pooling Tokens \\
         & TaobaoAd & q16 h64, Seq Pooling Tokens \\
         & TAAC & q16 h64, Seq Pooling Tokens \\
         & KuaiRand & q16 h128, Seq Pooling Tokens \\
\midrule
OneTrans & KuaiRec & h128, 4 layers, 4 heads, no causal mask \\
            & TaobaoAd & h64, 4 layers, 4 heads, no causal mask \\
            & TAAC & h64, 4 layers, 4 heads, no causal mask \\
            & KuaiRand & h128, 4 layers, 4 heads, no causal mask, batch 1024 \\
\midrule
MixFormer & KuaiRec & 4 heads, head dim 16, 4 layers, seq emb 64, sparse emb 32 \\
          & TaobaoAd & 4 heads, head dim 16, 4 layers, seq emb 32, sparse emb 32 \\
          & TAAC & 4 heads, head dim 16, 2 layers, seq emb 16, sparse emb 16 \\
          & KuaiRand & 4 heads, head dim 32, 4 layers, seq emb 64, sparse emb 32 \\
\bottomrule
\end{tabularx}
\caption{Selected baseline configurations per dataset for the four backbones in Table~\ref{tab:unified_results}.}
\label{tab:baseline_configs}
\end{table}

\subsection{KuaiRec Full-LENS Controlled vs Full-side Diagnostic}
\label{app:full_lens_robustness}

Table~\ref{tab:full_lens_robustness_app} contrasts Full LENS under two input scopes on dense KuaiRec: \emph{controlled} uses sequence and target only (no non-sequential side fields), while \emph{full-side} matches the main-table configuration. Full LENS is the strongest model under both scopes, but its full-side AUC is $0.0055$ lower than the controlled scope. The interpretation is that on a dense item vocabulary the controlled inputs already saturate the target-ID and sequence signals; injecting additional metadata through QueryGen can dilute the TCPB-conditioned position retrieval. This is consistent with the gate-only feature-path diagnostic in Section~\ref{sec:robustness}, where DIN exhibits the same density-regime behaviour more sharply. The same TCPB module remains beneficial under the main full-side LENS on the other three datasets, where the target-ID is less reliable on its own.

\begin{table}[t]
\centering
\small
\begin{tabular}{lrrr}
\toprule
\textbf{Model} & \textbf{Controlled} & \textbf{Full-side} & \textbf{$\Delta$} \\
\midrule
Full LENS & $0.8473 \pm 0.0027$ & $0.8418 \pm 0.0012$ & $-0.0055$ \\
\bottomrule
\end{tabular}
\caption{KuaiRec Full LENS AUC under controlled (no non-sequential fields) and full-side input scopes, q12 h64, seeds 42/123/456.}
\label{tab:full_lens_robustness_app}
\end{table}

\subsection{LENS Component Attribution Visualization}
\label{app:component_attribution_fig}

\begin{figure}[t]
  \centering
  \includegraphics[width=\linewidth]{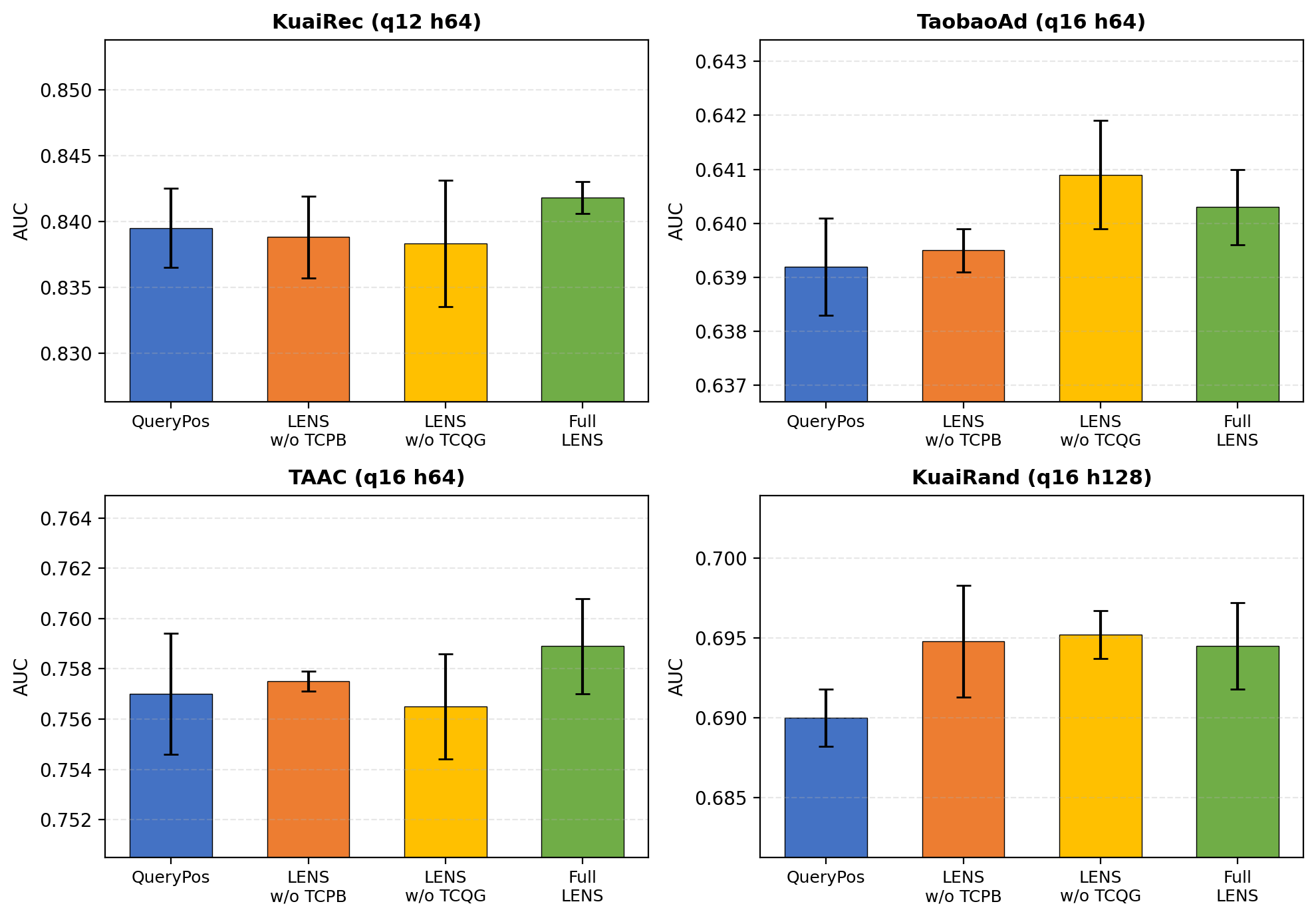}
  \caption{LENS component attribution on the HyFormer+QueryPos base (three seeds), one panel per dataset.}
  \label{fig:component_ablation_app}
\end{figure}

Figure~\ref{fig:component_ablation_app} visualizes Part IV of Table~\ref{tab:unified_results} as per-dataset bar charts. Starting from the HyFormer+QueryPos baseline, adding either TCQG or TCPB alone produces partial gains; combining them matches or recovers the best variant. TCPB is the dominant contributor on TaobaoAd and KuaiRand, while Full LENS is strongest on KuaiRec and TAAC.

\subsection{Full Position-Bias Visualization}
\label{app:full_position_bias}

\begin{figure}[t]
  \centering
  \includegraphics[width=\linewidth]{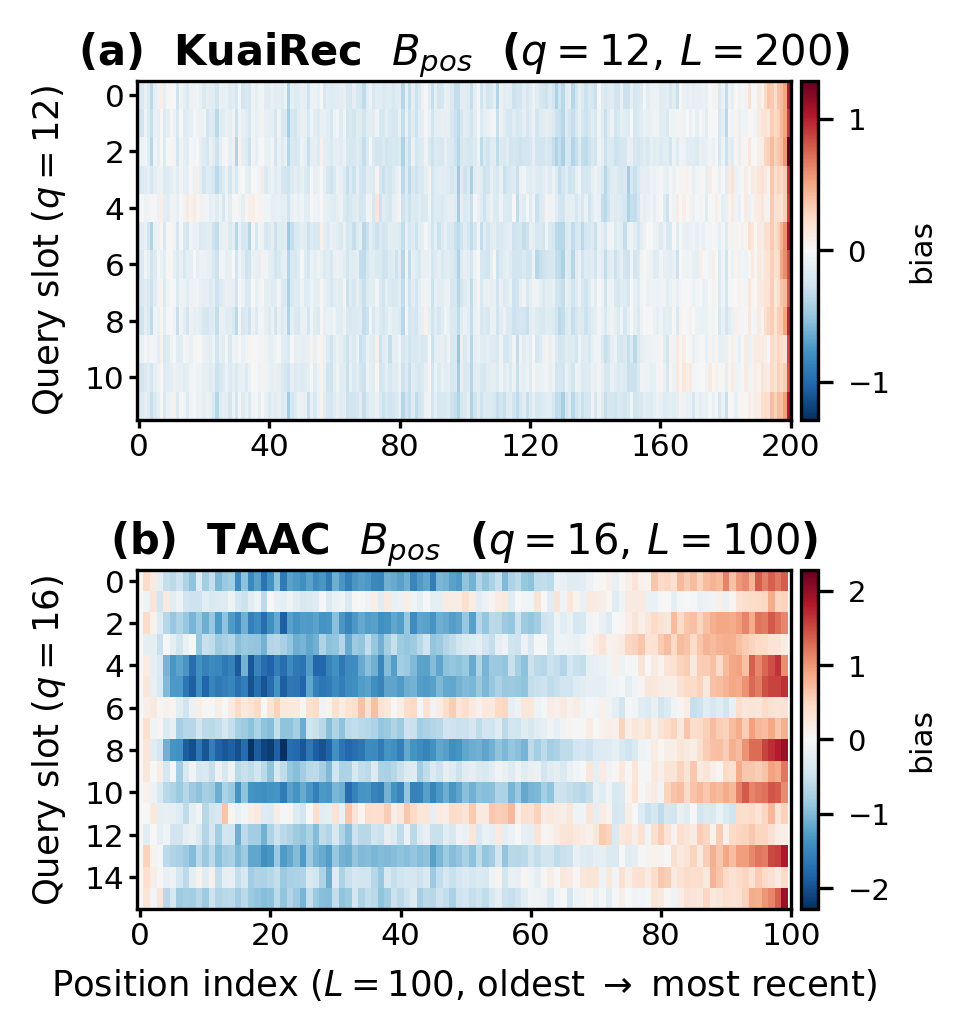}
  \caption{Learned Query-Specific Position Bias matrices: (a) KuaiRec ($q{=}12$, $L{=}200$); (b) TAAC ($q{=}16$, $L{=}100$). Color encodes bias values.}
  \label{fig:position_bias_full}
\end{figure}

Figure~\ref{fig:position_bias_full} shows the full learned Query-Specific Position Bias matrices on KuaiRec and TAAC. The two datasets behave differently. On KuaiRec, the 12 latent queries largely converge to a shared recent-position preference, so a single shared position curve already captures most of the position-aware signal. On TAAC, the 16 queries differentiate into more diverse temporal profiles, with some queries concentrated on recent positions and others retaining a broader historical window. This contrast is why the main text keeps only the TAAC view: the heterogeneity of TAAC queries is what makes target-conditioned query activation a meaningful complement to a static position prior.

\subsection{DIN Feature-Path Per-Seed Details}
\label{app:din_diagnostics}

This subsection expands the feature-path analysis from Section~\ref{sec:robustness}. The key diagnostic on KuaiRec is that DIN recovers its native accuracy when the sparse branch parameters are frozen at zero, but drops when the branch is trainable; the sparse-feature MLP branch creates a competing gradient path that degrades the native target-attention mechanism. Table~\ref{tab:din_per_seed} reports the per-seed AUCs underlying this conclusion: native DIN is the best variant on every seed, and adding full-side metadata is consistently worst. On TAAC, by contrast, the two-ID sparse path alone does not help (native and +two-ID are tied at $\sim$0.716), but full-side metadata improves DIN to $0.748$. The feature-path effect is therefore density-regime dependent rather than architecture-specific.

\begin{table}[t]
\centering
\small
\begin{tabular}{lrrr}
\toprule
\textbf{DIN Variant} & \textbf{seed42} & \textbf{seed123} & \textbf{seed456} \\
\midrule
Native no-sparse & 0.8351 & 0.8412 & 0.8383 \\
+ two-ID sparse & 0.8282 & 0.8369 & 0.8328 \\
+ full-side metadata & 0.8239 & 0.8317 & 0.8288 \\
\bottomrule
\end{tabular}
\caption{KuaiRec DIN feature-path per-seed AUC, positive-history protocol.}
\label{tab:din_per_seed}
\end{table}

\section{HyFormer+ Switch Ablation: Per-Seed Details}
\label{app:switch_ablation}

This appendix provides per-seed values supporting the HyFormer switch ablation in Part I of Table~\ref{tab:unified_results}. The three-seed mean and standard deviation already appear in the main table; the values below are the underlying per-seed AUCs.

\begin{table*}[!t]
\centering
\small
\resizebox{\textwidth}{!}{%
\begin{tabular}{llrrrr}
\toprule
\textbf{Dataset} & \textbf{Switch} & \textbf{seed42} & \textbf{seed123} & \textbf{seed456} & \textbf{Mean $\pm$ Std} \\
\midrule
KuaiRec q12 h64 & HyFormer (no switches) & 0.8207 & 0.8273 & 0.8299 & $0.8260 \pm 0.0047$ \\
KuaiRec q12 h64 & HyFormer + Seq Pooling Tokens & 0.8287 & 0.8358 & 0.8339 & $\mathbf{0.8328 \pm 0.0037}$ \\
KuaiRec q12 h64 & HyFormer + NS Tokens in Query Boosting & 0.8220 & 0.8275 & 0.8263 & $0.8253 \pm 0.0029$ \\
KuaiRec q12 h64 & HyFormer + Per-query FFN & 0.8253 & 0.8329 & 0.8299 & $0.8294 \pm 0.0038$ \\
KuaiRec q12 h64 & HyFormer + Full Switches & 0.8227 & 0.8334 & 0.8314 & $0.8292 \pm 0.0057$ \\
\midrule
TAAC q16 h64 & HyFormer (no switches) & 0.7564 & 0.7545 & 0.7571 & $0.7560 \pm 0.0013$ \\
TAAC q16 h64 & HyFormer + Seq Pooling Tokens & 0.7530 & 0.7558 & 0.7572 & $0.7553 \pm 0.0021$ \\
TAAC q16 h64 & HyFormer + NS Tokens in Query Boosting & 0.7568 & 0.7591 & 0.7533 & $0.7564 \pm 0.0029$ \\
TAAC q16 h64 & HyFormer + Per-query FFN & 0.7565 & 0.7596 & 0.7573 & $\mathbf{0.7578 \pm 0.0016}$ \\
TAAC q16 h64 & HyFormer + Full Switches & 0.7499 & 0.7466 & 0.7548 & $0.7504 \pm 0.0041$ \\
\midrule
TaobaoAd q16 h64 & HyFormer (no switches) & 0.6402 & 0.6406 & 0.6403 & $\mathbf{0.6404 \pm 0.0002}$ \\
TaobaoAd q16 h64 & HyFormer + Seq Pooling Tokens & 0.6404 & 0.6387 & 0.6390 & $0.6394 \pm 0.0009$ \\
TaobaoAd q16 h64 & HyFormer + NS Tokens in Query Boosting & 0.6397 & 0.6393 & 0.6391 & $0.6394 \pm 0.0003$ \\
TaobaoAd q16 h64 & HyFormer + Per-query FFN & 0.6404 & 0.6387 & 0.6385 & $0.6392 \pm 0.0010$ \\
TaobaoAd q16 h64 & HyFormer + Full Switches & 0.6394 & 0.6393 & 0.6399 & $0.6395 \pm 0.0003$ \\
\midrule
KuaiRand typed q16 h128 & HyFormer (no switches) & 0.6733 & 0.6645 & 0.6715 & $0.6698 \pm 0.0046$ \\
KuaiRand typed q16 h128 & HyFormer + Seq Pooling Tokens & 0.6671 & 0.6700 & 0.6665 & $0.6679 \pm 0.0019$ \\
KuaiRand typed q16 h128 & HyFormer + NS Tokens in Query Boosting & 0.6640 & 0.6681 & 0.6480 & $0.6600 \pm 0.0106$ \\
KuaiRand typed q16 h128 & HyFormer + Per-query FFN & 0.6706 & 0.6705 & 0.6628 & $0.6680 \pm 0.0045$ \\
KuaiRand typed q16 h128 & HyFormer + Full Switches & 0.6800 & 0.6751 & 0.6690 & $\mathbf{0.6747 \pm 0.0055}$ \\
\bottomrule
\end{tabular}
}%
\caption{Seed-level AUC values for the HyFormer+ switch ablation summarized in Part I of Table~\ref{tab:unified_results}. Standard deviations use sample std ($ddof=1$). KuaiRand's ``NS Tokens in Query Boosting'' variant has the highest variance ($\mathrm{std}=0.0106$), suggesting instability when non-sequential tokens enter Boosting on this extreme-sparse dataset; this is the only switch combination flagged for downstream caution.}
\label{tab:switch_ablation_raw}
\end{table*}

\end{document}